\documentclass[aps,prd,reprint,twocolumn,superscriptaddress,showpacs]{revtex4-2}

\usepackage{graphicx}
\usepackage{mathrsfs}
\usepackage{bm}
\usepackage{amsmath}
\usepackage{dcolumn}
\usepackage{epstopdf}
\usepackage{dsfont}
\usepackage{amssymb}
\usepackage{tabularx}
\usepackage{array}
\usepackage{float}
\usepackage{color}
\usepackage{epstopdf}
\usepackage{mathrsfs}
\usepackage[colorlinks, linkcolor=blue,anchorcolor=blue,citecolor=blue,urlcolor=blue]{hyperref}
\usepackage{multirow}

\begin{document}
\title{Chiral Topological Phononic Quasiparticles in Enantiomeric Crystals SrSi$_2$ and BaSi$_2$}
\author{Yong-Kun Wang}
\affiliation{School of Physics, Northwest University, Xi'an 710127, China}
\affiliation{Shaanxi Key Laboratory for Theoretical Physics Frontiers, Xi'an 710127, China}

\author{An-Dong Fan}
\affiliation{School of Physics, Northwest University, Xi'an 710127, China}
\affiliation{Shaanxi Key Laboratory for Theoretical Physics Frontiers, Xi'an 710127, China}

\author{Jin-Yang Li}
\affiliation{School of Physics, Northwest University, Xi'an 710127, China}
\affiliation{Shaanxi Key Laboratory for Theoretical Physics Frontiers, Xi'an 710127, China}

\author{Huaqing Huang}
\email{huaqing.huang@pku.edu.cn}
\affiliation{School of Physics, Peking University, Beijing 100871, China}

\author{Si Li}
\email{sili@nwu.edu.cn}
\affiliation{School of Physics, Northwest University, Xi'an 710127, China}
\affiliation{Shaanxi Key Laboratory for Theoretical Physics Frontiers, Xi'an 710127, China}

\begin{abstract}
Chiral crystals have recently garnered significant interest in condensed matter physics due to their unique electronic and optical properties. In this paper, we explore the connection between the chirality of crystal structures and the chirality of topological quasiparticles. We specifically predict and analyze several chiral enantiomeric materials, such as SrSi$_2$ and BaSi$_2$, which crystallize in the chiral space groups $P4{_3}32$ and $P4{_1}32$. Based on first-principles calculations and theoretical analysis, we reveal that the phonon spectra of these materials host various topological phononic quasiparticles, including charge-2 triple points, charge-2 Dirac points, charge-2 Weyl points, and charge-1 Weyl points. Our paper shows that in these enantiomeric materials, the opposite chirality of the crystal structure results in topological quasiparticles with opposite chiral topological charges and distinct topological surface states. Our paper elucidates the intrinsic relationship between the chirality of crystal structures and the chirality of topological quasiparticles, providing promising theoretical guidance and material platform for investigating the physical properties of chiral crystals.
\end{abstract}
\maketitle
\section{Introduction}
Chirality is a geometric property that renders an object distinguishable from its mirror image, making it impossible to superimpose the object onto its reflection. Such objects are referred to as chiral, and each chiral object can exist in two distinct configurations known as enantiomers. This property is widespread across various scientific fields, including chemistry, biology, and physics, and is particularly significant due to its role in creating unique characteristics within materials~\cite{fecher2022chirality,felser2023topology,zhang2024structural,bousquet2024structural}. In condensed matter physics, chiral materials exhibit remarkable properties stemming from the absence of certain symmetries, such as mirror, inversion, or rotoinversion symmetries. These symmetry-breaking characteristics can give rise to novel phenomena and have potential applications across electronics, optics, and materials science~\cite{yang2021chiral,wang2024chiral,yan2024structural}. Examples of such phenomena include skyrmions~\cite{bogdanov1994thermodynamically}, non-local and non-reciprocal electron transport~\cite{rikken2001electrical,yoda2015current}, optical activity and magnetochiral dichroism~\cite{fasman2013circular}, unusual superconductivity~\cite{carnicom2018tarh2b2}, ferroelectricity~\cite{mitamura2014spin,hu2020chiral}, and spintronics~\cite{yang2020spintronics}. Recent research has focused on the topological properties of chiral crystals~\cite{shekhar2018chirality,chang2018topological}, leading to the discovery of various novel topological chiral fermions, such as Kramers-Weyl fermions and higher-fold degenerate fermions~\cite{chang2018topological,sanchez2019topological,li2019chiral,xie2021kramers,bradlyn2016beyond,chang2017unconventional,tang2017multiple}.

Topological quasiparticles in condensed matter physics, which typically emerge around band degeneracy points in the band structure, have exhibited novel behaviors that are distinct from conventional fermions \cite{armitage2018weyl,zhang2019catalogue,vergniory2019complete,tang2019comprehensive,xu2020high,lv2021experimental,vergniory2022all,yu2022encyclopedia}. 
For instance, Weyl and Dirac points, among the most well-known topological quasiparticles, feature isolated twofold and fourfold degenerate linear band-crossing points, respectively, where the electronic excitations resemble relativistic Weyl and Dirac fermions~\cite{murakami2007phase,wan2011topological,armitage2018weyl,young2012dirac,wang2012dirac,wang2013three}. This allows for the simulation of intriguing high-energy physics phenomena within condensed matter systems. However, unlike high-energy physics, condensed matter systems are constrained by space group (SG) symmetries rather than Poincaré symmetry, leading to a much richer variety of unconventional topological quasiparticles. These include nodal lines~\cite{burkov2011topological,weng2015topological,Chen2015,Mullen2015,Fang2015,Yu2015,Kim2015,Li2016,Bian2016,Huang2016,Yu2017,Li2017,Li2018nonsymmorphic}, nodal surfaces~\cite{zhong2016towards,liang2016node,bzduvsek2017robust,wu2018nodal}, high-fold fermions (three-, six-, and eight-fold degenerate fermions)\cite{weng2016topological,zhu2016triple,bradlyn2016beyond}, and higher-order fermions with quadratic and cubic dispersions\cite{fang2012multi,yu2019quadratic}. Topological quasiparticles typically exhibit fascinating physical properties, such as Weyl points acting as monopoles of Berry curvature in momentum space and carrying nonzero Chern numbers (topological chiral charges) denoted as $C=\pm 1$, indicating the chirality of Weyl fermions. These nonzero chiral charges lead to various unusual physical phenomena, including negative magnetoresistance effects~\cite{son2013chiral,huang2015observation} and helical surface states~\cite{fang2016topological}.

While topological quasiparticles were initially explored in electronic systems, subsequent research has extended these concepts to bosonic and classical systems, including phonon systems~\cite{zhang2010topological,zhang2015chiral,susstrunk2016classification,liu2017pseudospins,ji2017topological,liu2020topological}.
 Phonons, the fundamental bosonic excitations in crystalline materials, represent the collective vibrations of the atomic lattice. They play a vital role in determining a material's thermal properties and interact closely with other quasiparticles or collective excitations, such as electrons, photons, and magnons. Numerous materials exhibiting topological phonons with nodal points, nodal lines, and nodal surfaces have been theoretically predicted~\cite{stenull2016topological,susstrunk2016classification,	liu2017pseudospins,ji2017topological,singh2018topological,zhang2018double,miao2018observation,zhang2019phononic,liu2020topological,li2021computation,jin2021tunable,xie2021sixfold,zhang2023weyl,xu2024catalog}, with several of these predictions successfully validated through experimentation~\cite{miao2018observation,zhang2019phononic,li2021observation,zhang2023weyl}.

\begin{figure}[htb]
	\includegraphics[width=8.5cm]{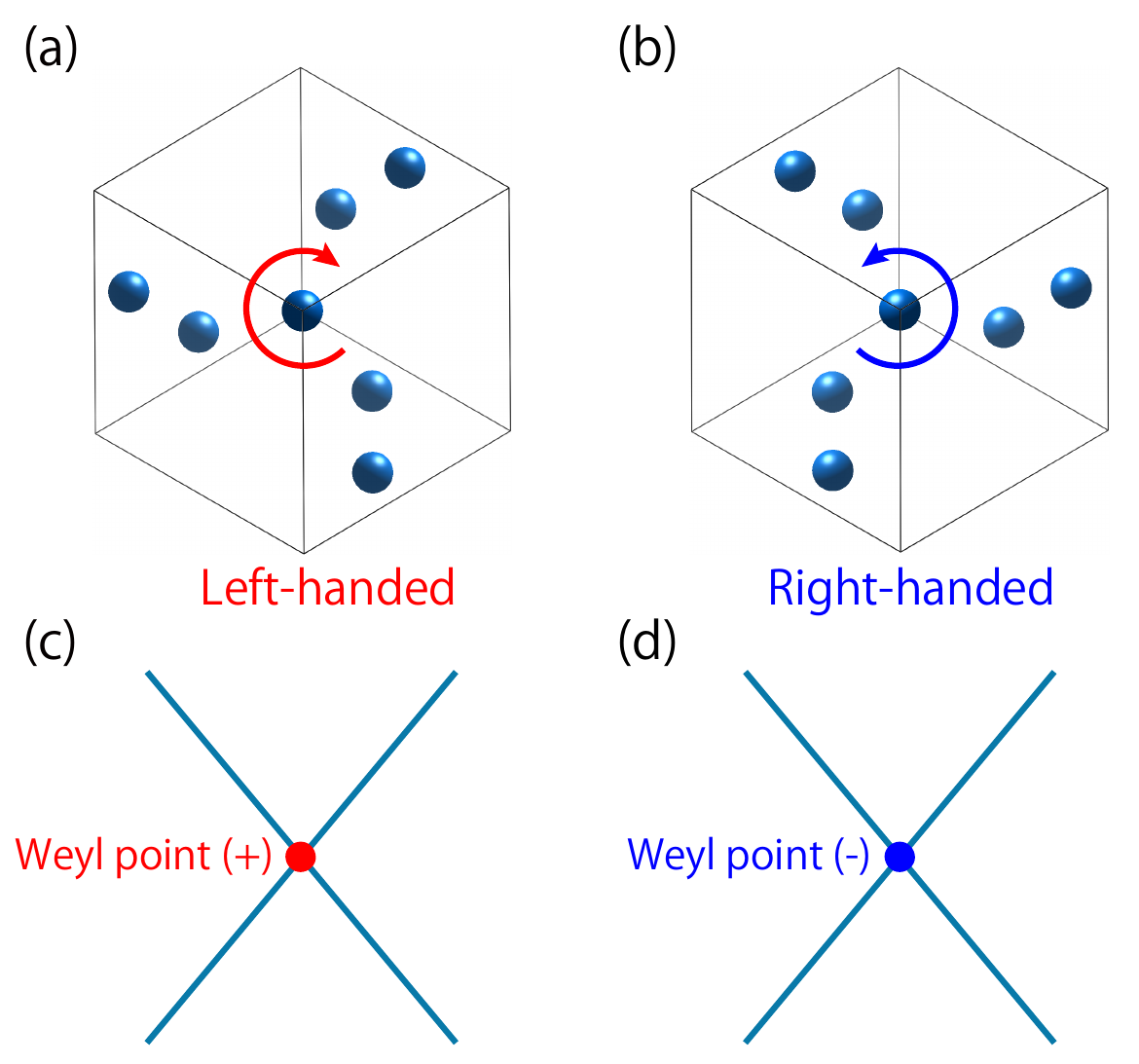}
	\caption{Schematic diagrams of the two enantiomers of chiral crystal materials: (a) left-handed chiral crystal, and (b) right-handed chiral crystal. Schematic diagrams of Weyl points formed in chiral crystal materials: (c) Weyl points with positive topological charge in left-handed chiral crystals, and (d) Weyl points with negative topological charge in right-handed chiral crystals.
	\label{fig1}}
\end{figure}

Chiral crystals, which exhibit distinct chiralities and form enantiomers, can influence the chirality of topological phononic quasiparticles, leading to intriguing questions about the relationship between the chirality of crystal structure and chiral topological charges of phononic quasiparticles. In this work, we investigate this relationship and reveal that chiral crystals can induce chiral topological phononic quasiparticles, with enantiomers possessing opposite chirality yielding opposite chiral charges (For example, if there is a Weyl point with a topological charge of $+1$ in the momentum space of a left-handed chiral crystal, then there should be a Weyl point with a topological charge of $-1$ at the same momentum space location in the right-handed chiral crystal, as illustrated in Fig.~\ref{fig1}). Our first-principles calculations and theoretical analysis demonstrate that various topological quasiparticles, including charge-2 triple points (C-2 TPs), charge-2 Dirac points (C-2 DPs), charge-2 Weyl points (C-2 WPs), and charge-1 Weyl points (C-1 WPs), can be realized in the phonon spectra of chiral crystal materials such as SrSi$_2$ and BaSi$_2$. We further show that the opposite chirality of these enantiomeric structures leads to topological phononic quasiparticles with opposite chiral topological charges and distinct topological surface states.

\section{Computation METHODS}
The first-principles calculations were conducted using the density functional theory (DFT) within the Vienna \emph{ab initio} simulation package~\cite{kresse1994,kresse1996,blochl1994projector}, employing the projector augmented wave method~\cite{blochl1994projector}. The exchange-correlation functional was modeled with the generalized gradient approximation (GGA) using the Perdew-Burke-Ernzerhof (PBE) parameterization~\cite{PBE}. A cutoff energy of 500 eV was applied, and the Brillouin zone (BZ) was sampled with a $\Gamma$-centered $k$ mesh of size $8\times 8\times 8$. Convergence criteria were set to $10^{-8}$ eV for energy and $10^{-3}$ eV/\AA\ for forces. The force constants and phonon spectra were obtained using the density functional perturbation theory (DFPT) with a $2 \times 2 \times 2$ supercell and a $3\times 3\times 3$ $k$-mesh, in combination with the Phonopy package~\cite{togo2015first}. Irreducible representations of phonon spectra were determined using the PhononIrep package~\cite{zhang2022phonon}. To calculate Chern numbers and surface states, the Green’s function method implemented in the WannierTools package was utilized~\cite{sancho1984quick,sancho1985highly,wu2018wanniertools}.

\section{Chiral Crystal Structures}
\begin{figure}[htb]
	\includegraphics[width=8.5cm]{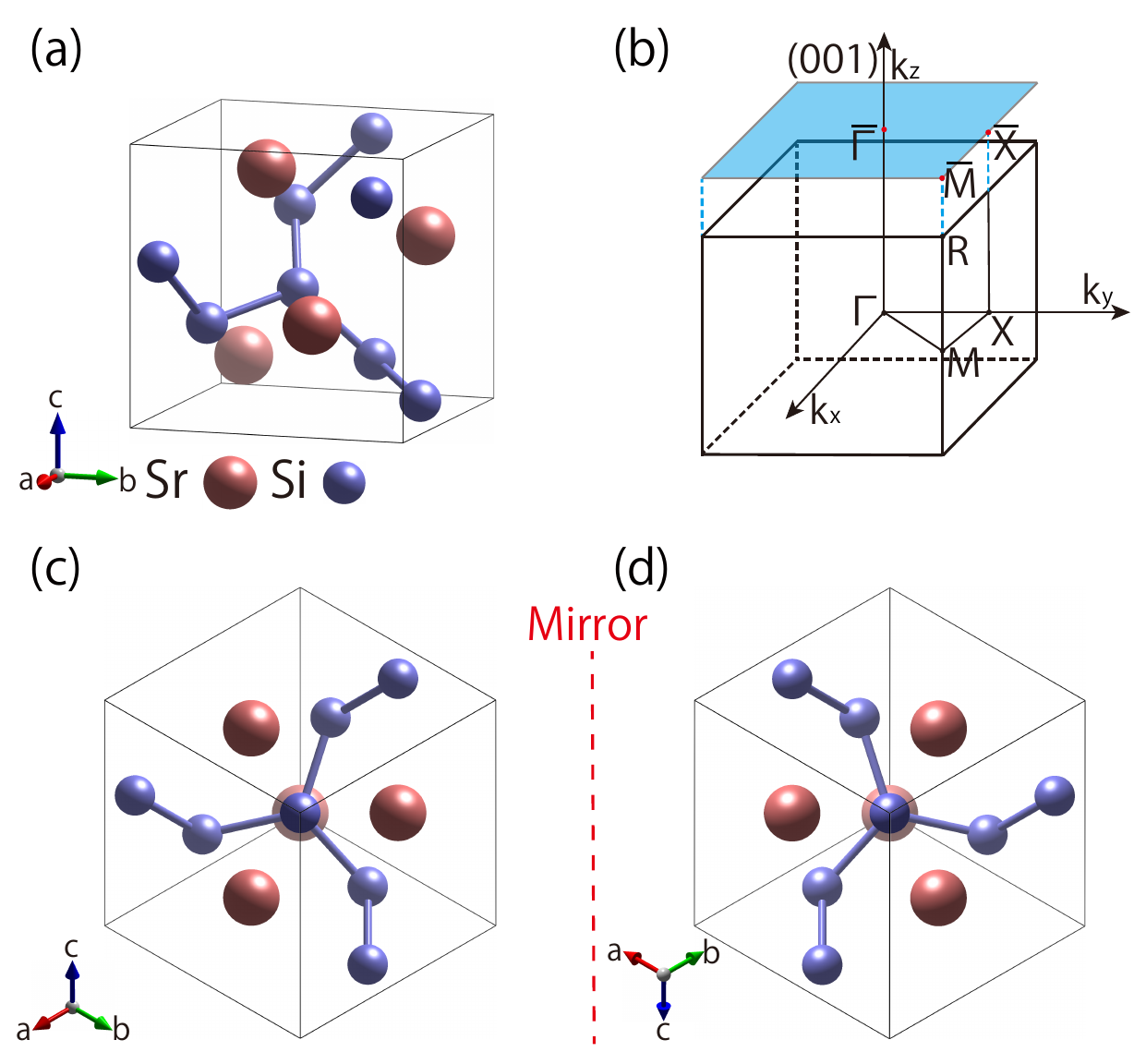}
	\caption{(a) Side view of the crystal structure of SrSi$_2$ with space group No. 212. (b) The bulk Brillouin zone (BZ) and the projected surface BZ of the (001) plane, with high-symmetry points labeled. (c) The view of SrSi$_2$ with space group No. 212 along the [111] direction, and (d) the view of SrSi$_2$ with space group No. 213 along the [$\bar1\bar1\bar1$] direction. The two enantiomers are related by mirror symmetry. 
		\label{fig2}}
\end{figure}

Due to the similar crystal structures and phonon energy bands of BaSi$_2$ and SrSi$_2$ materials, we primarily present the results of SrSi$_2$ in the main text, while the results of BaSi$_2$ are shown in the Supplemental Material (SM)~\cite{SM}.
The single-crystal SrSi$_2$ material has been experimentally synthesized and it crystallizes in a cubic Bravais lattice with the chiral SG No. 212 ($P4{_3}32$) or No. 213 ($P4{_1}32$)~\cite{evers1978transformation}. The generating elements for SG No. 212 are $\{C_{2z}|\frac{1}{2},0,\frac{1}{2}\}$, $\{C_{2x}|\frac{1}{2},\frac{1}{2},0\}$, $\{C_{2,110}|\frac{1}{4},\frac{3}{4},\frac{3}{4}\}$,  $\{C^{-1}_{3,111}|0,0,0\}$ and for SG No. 213 are $\{C_{2z}|\frac{1}{2},0,\frac{1}{2}\}$, $\{C_{2x}|\frac{1}{2},\frac{1}{2},0\}$, $\{C_{2,110}|\frac{3}{4},\frac{1}{4},\frac{1}{4}\}$,  $\{C^{-1}_{3,111}|0,0,0\}$. The crystal structure and the bulk and (001) surface Brillouin zones (BZ) of SrSi$_2$ are shown in Figs.~\ref{fig2} (a) and (b), respectively. From Fig.~\ref{fig2} (a), one can observe each unit cell
contains four Sr atoms and eight Si atoms, in which the Si atoms form three-dimensional three-connected nets. In our calculation, the structure is
fully optimized (see SM~\cite{SM} for computational details). The optimized lattice constant is a = 6.564 \AA, which is in good agreement with the experimental value (a = 6.535  \AA)~\cite{evers1978transformation}, and is almost consistent with other theoretical calculations~\cite{huang2016new,huang2020three}. The calculated Wyckoff positions of Sr and Si atoms are $4a$ (0.125, 0.125, 0.125) and $8c$ (0.4224, 0.4224, 0.4224) for SG No. 212, and are $4b$ (0.875, 0.875, 0.875) and $8c$ (0.5776, 0.5776, 0.5776) for SG No. 213. The atoms coordinates are measured in units of the respective lattice parameters. It should be noted that the structures with SG Nos. 212 and 213 are chiral enantiomers, as shown in Fig.~\ref{fig2} (c) and (d). 

\section{Multiple Types of Chiral Quasiparticles.}
\begin{figure*}[htb]
	\includegraphics[width=18cm]{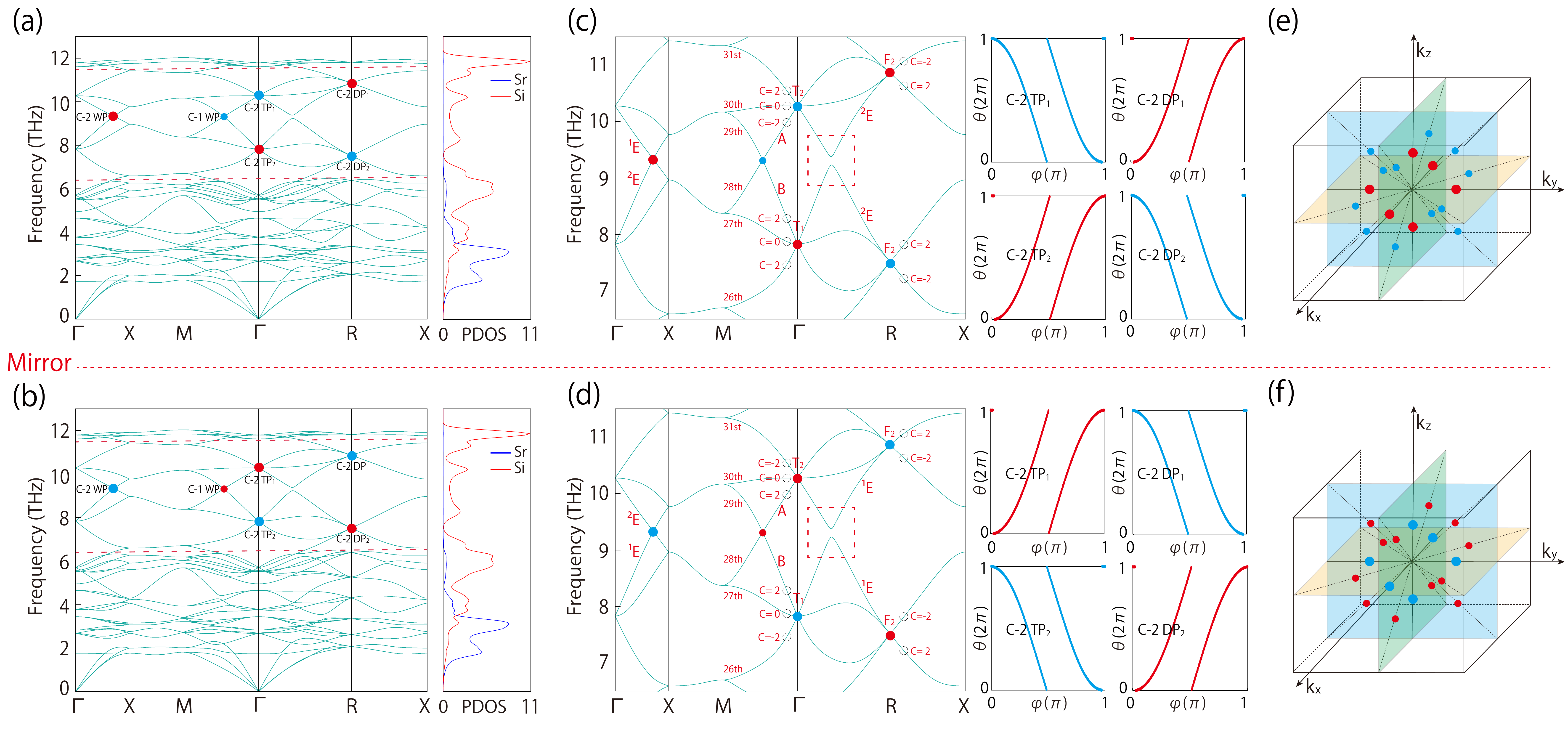}
	\caption{Panels (a) and (b) are the phonon dispersion and projected density of states (PDOS) of SrSi$_2$ with SG Nos. 212 and 213, respectively. (c) and (d) are the zoom-in of the phonon spectrum for the six phonon bands (the 26th to the 31st bands) that locate between the two dashed red lines in (a) and (b), respectively. The irreducible representations (IRRs), Chern numbers, and the evolutions of WCCs for C-2 TP and C-2 DP are given. (e) and (f) distribution of Weyl points in reciprocal space for SrSi$_2$ with SG Nos. 212 and 213, respectively. Large spheres represent C-2 Weyl points and small spheres represent C-1 Weyl points. Red spheres represent positive topological charge, while green spheres represent negative topological charge.
		\label{fig3}}
\end{figure*}
The phonon spectrum and the projected density of states (PDOS) of SrSi$_2$ material with SG Nos. 212 and 213, obtained from first-principles calculations, is shown in Figs.~\ref{fig3} (a) and (b), respectively. As seen in Figs.~\ref{fig3} (a) and (b), six phonon bands (the 26th to 31st bands) between 6.5 and 11.5 THz are connected, as indicated by the two dashed red lines. The zoomed-in images of these phonon bands between the dashed red lines are displayed in Figs.~\ref{fig3} (c) and (d), along with the calculated irreducible representations (IRRs) for the phonon bands. As illustrated in Figs.~\ref{fig3} (c) and (d), the six entangled phonon bands form multiple types of topological quasiparticles, such as threefold degenerate points at the $\Gamma$ point, fourfold degenerate Dirac points at the $R$ point, hourglass dispersions along the $\Gamma$-X and M-$\Gamma$ paths, and twofold degeneracies on the X-M path. Through theoretical analysis and Chern number calculations, we find that the threefold degenerate points at the $\Gamma$ point correspond to C-2 TPs (also known as spin-1 Weyl points), the Dirac points at the $R$ point correspond to C-2 DPs, and the neck points of the hourglass dispersion form C-2 WPs on the $\Gamma$-X path and C-1 WPs on the $\Gamma$-M path. From the PDOS, one can observe that the phonon bands of the topological quasiparticles are mainly formed by the Si atoms. Additionally, all topological quasiparticles exhibit opposite chirality for the two enantiomers with SG Nos. 212 and 213. In the following sections, we will discuss these topological quasiparticles in detail.

We first analyze the threefold degenerate points at the $\Gamma$ point. As depicted in Figs.~\ref{fig3}(c) and (d), two threefold degenerate points can be observed at the $\Gamma$ point, with their associated three-dimensional irreducible representations (IRRs) being $T_1$ and $T_2$, respectively. The calculated Chern numbers for the three bands forming these degenerate points are $\pm 2$ and 0, as shown in Figs.~\ref{fig3}(c) and (d). These threefold degenerate points at the $\Gamma$ point are also referred to as spin-1 Weyl points~\cite{zhang2018double}, which can be viewed as a monopole with a topological charge of 2. The threefold degenerate point with topological charge 2 at the $\Gamma$ point is also called a charge-2 triple point (C-2 TP). It is important to note that the topological charge of the C-2 TPs for the two enantiomers has opposite chirality, as illustrated in Figs.~\ref{fig3}(c) and (d) (also refer to Table~\ref{table1}.). The detailed symmetry analysis for the C-2 TP at the $\Gamma$ point is given in SM~\cite{SM}. To further characterize the C-2 TP, we construct a low-energy effective $k \cdot p$ model based on symmetry requirements(see the SM~\cite{SM}), and the model at the $\Gamma$ point, up to the first order in $\bf{k}$, is as follows:

\begin{equation}
	H_{TP}(k)=a \cdot\left(\begin{array}{ccc}
		0 & i k_z & -i k_y \\
		-i k_z & 0 & i k_x \\
		i k_y & -i k_x & 0
	\end{array}\right)=a \cdot \bf{k} \cdot \bf{S},
\end{equation}
where $a$ is a real constant, $S_i$ are the spin-1 matrix representations of the rotation generators. This model is just the spin-1 Weyl Hamiltonian.  

We then explore the fourfold degenerate Dirac points at the $R$ point in the BZ corner.
As shown in Figs.~\ref{fig3}(c) and (d), there are two fourfold degenerate Dirac points at the $R$ point, both corresponding to the four-dimensional irreducible representations (IRRs) $F_2$. The calculated Chern numbers of the bands around the Dirac points are $\pm 2$, indicating that they are charge-2 Dirac points (C-2 DPs). We find that the topological charges of the C-2 DPs for the two enantiomers exhibit opposite chirality (see Table~\ref{table1}.). The symmetry analysis of the C-2 DP is presented in the SM~\cite{SM}. Notably, the Dirac point at the $R$ point is independent of the materials and is solely determined by the crystal space group symmetry. Therefore, all the phonon bands at the $R$ point are fourfold degenerate. Additional, one can observe that the Chern numbers of the C-2 TPs and C-2 DPs are opposite [see Figs.~\ref{fig3}(c) and (d) and Table~\ref{table1}.], and they appear in pairs in the BZ. Therefore, the total Chern number is zero, consistent with the no-go theorem~\cite{nielsen1981absence}. We also construct a low-energy effective $k \cdot p$ model for the C-2 DP (see the SM~\cite{SM}), and the model at the $R$ point, up to the first order in $\bf{k}$, is as follows:
	\begin{equation}
		H_{DP}(k)=\left(
		\begin{array}{cccc}
			\xi_1 & \xi_5 & \xi_3 & \xi_4 \\
			\xi_5^{\dagger} & -\xi_1 & \xi_4 & \xi_3 \\
			\xi_3^{\dagger} & \xi_4^{\dagger} & \xi_2 & \xi_6 \\
			\xi_4^{\dagger} & \xi_3^{\dagger} & \xi_6^{\dagger}	& -\xi_2\\
		\end{array}
		\right)
	\end{equation}
	where 
	\begin{equation}
		\resizebox{1.0\hsize}{!}
		{$\begin{split}
				&\xi_1=\frac{\sqrt{2}}{4} \left(c_{1}-2 c_{2}\right) (k_x-k_y),~ \xi_2=\frac{\sqrt{2}}{4} \left(c_{1}+2 c_{2}\right) (k_x-k_y)\\
				&\xi_3=-i \frac{1}{2} \sqrt{\frac{3}{2}} c_{1} (k_x-k_y),~ \xi_4=-\frac{1}{2} \sqrt{\frac{3}{2}} c_{1}(k_x+k_y)\\
				&\xi_5= -\frac{c_{1}}{4} \left(\sqrt{2} ik_x+\sqrt{2} ik_y+4 k_z\right)-2\sqrt{2}c_{2} \left(k_x+k_y+\sqrt{2}ik_z\right)\\
				&\xi_6= \frac{c_{1}}{4} \left(\sqrt{2} ik_x+\sqrt{2} ik_y-4 k_z\right)+ \frac{c_{2}}{2}\left(\sqrt{2}ik_x+\sqrt{2}ik_y+2 k_z\right)
			\end{split}$}
	\end{equation}
	and $c_1,c_2$ are real parameters. After a unitary transformation,
	\begin{equation}
		H_D(k)=b \left(\begin{array}{cc}
			\bf{k} \cdot \bf{\sigma} & 0 \\
			0 & \bf{k} \cdot \bf{\sigma} \\
		\end{array}\right)
	\end{equation}
	where $b$ is a nonzero constant. This model is the direct sum of two identical spin-1/2 Weyl points, and is also referred to as C-2 DP.
	
\begin{table}[t]
	\caption{\label{table1}The corresponding frequencies (THz), positions, and topological charges of the topological quasiparticles C-2 TP and C-2 DP in SrSi$_2$ for the two enantiomers.}
	\begin{ruledtabular}
		\begin{tabular}{ccccc}
			SG   & Quasiparticles & Frequency & Position & Charge\\
			\hline 	
			\multirow{4}{*}{$P4_{3}32$}  & C-2 TP$_1$    & 10.27 & $\Gamma$ &  $-2$ \\
			& C-2 DP$_1$    & 10.86 & R        &  $+2$ \\
			& C-2 TP$_2$    &  7.83  & $\Gamma$ &  $+2$  \\
			& C-2 DP$_2$    &  7.51  & R        &  $-2$ \\
			\hline 	
			\multirow{4}{*}{$P4_{1}32$}  & C-2 TP$_1$    & 10.27 & $\Gamma$ &  $+2$  \\
			& C-2 DP$_1$    & 10.86 & R        &  $-2$  \\
			& C-2 TP$_2$    & 7.83  & $\Gamma$ &  $-2$  \\
			& C-2 DP$_2$    & 7.51  & R        &  $+2$ \\ 
		\end{tabular}
	\end{ruledtabular}
\end{table}

Next, we discuss the band crossings along the $\Gamma$-X and M-$\Gamma$ paths. As shown in Figs.~\ref{fig3}(c) and (d), the four phonon bands (the 27th to the 30th) exhibit hourglass dispersions as they cross along these paths. Note that there is a small gap between the 28th and 29th bands along the $\Gamma$-R path. We calculated the IRRs for these bands along the $\Gamma$-X, M-$\Gamma$, and $\Gamma$-R paths, as depicted in Figs.~\ref{fig3}(c) and (d). The 28th and 29th bands have different representations: $^1E$ and $^2E$ along the $\Gamma$-X path, and $B$ and $A$ along the M-$\Gamma$ path, but share the same representation (either $^2E$ or $^1E$) along the $\Gamma$-R path. Consequently, the 28th and 29th bands form band crossings along the $\Gamma$-X and M-$\Gamma$ paths. These band crossing points are, in fact, Weyl points with different topological charges.
Our first-principles calculations and theoretical analysis reveal that the Weyl point on the $\Gamma$-X path has quadratic dispersion along the $k_y/k_z$ axes and linear dispersion along the $k_x$ axis, with a topological charge of 2, hence referred to as a charge-2 Weyl point (C-2 WP). Meanwhile, the Weyl point on the M-$\Gamma$ path is a conventional Weyl point with linear dispersion along all axes and a topological charge of 1, referred to as a charge-1 Weyl point (C-1 WP). The two-dimensional band structures around the C-2 WP and C-1 WP, along with the Chern number calculations, are detailed in the SM~\cite{SM}.
Considering the $C_2$ and $C_4$ rotational symmetries in SrSi$_2$, there should be six C-2 WPs along the $C_4$ axes ($k_x$, $k_y$, and $k_z$) and twelve C-1 WPs along the $C_2$ axes, as illustrated in Figs.~\ref{fig3}(e) and (f). The topological charges of the C-2 WPs and C-1 WPs for the two enantiomers also exhibit opposite chirality [see SM~\cite{SM}, Table~\ref{table2}, and Figs.~\ref{fig3}(e) and (f)]. It is noteworthy that the Weyl state in SrSi$_2$ with space group $P4{_1}32$, induced by external pressure, has been experimentally observed~\cite{yao2024high}.
\begin{table}[t]
	\caption{\label{table2} The corresponding lattice parameters, positions, and topological charges of the  C-2 WP and C-1 WP in SrSi$_2$ for the two enantiomers.}
	\begin{ruledtabular}
		\begin{tabular}{ccccc}
			SG   & Lattice parameters & WPs & Position & Charge\\
			\hline 	
			\multirow{2}{*}{$P4_{3}32$}  & \multirow{2}{*}{6.564 \AA} & C-2 WP  & (0.357, 0,  0) &  $+2$ \\
			&    & C-1 WP  & (0.231,  0.231,  0) &  $-1$ \\

			\hline 	
			\multirow{2}{*}{$P4_{1}32$}  &  \multirow{2}{*}{6.564 \AA}   & C-2 WP  & (0.357, 0,  0) &  $-2$  \\
			&  & C-1 WP  & (0.231,  0.231,  0) &  $+1$  \\
		\end{tabular}
	\end{ruledtabular}
\end{table}
Finally, Figs.~\ref{fig3}(a) and (b) clearly shows that all the phonon bands along the X-M and R-X paths are twofold degenerate. Symmetry analysis and first-principles calculations reveal that all the bands on the $k_x = \pi$ plane are twofold degenerate, forming a nodal surface. Additionally, we identified two other nodal surfaces on the $k_y = \pi$ and $k_z = \pi$ planes, respectively, resulting in a total of three nodal surfaces at the BZ boundary. These nodal surfaces are protected by the combination of time-reversal symmetry $\mathcal{T}$ and two-fold screw rotation symmetry $\widetilde{\mathcal{C}}_{2i}$ (where $i = x, y, z$). Since in the absence of spin-orbit coupling (SOC), we have $(T\widetilde{\mathcal{C}}_{2i})^{2} = e^{-ik_{i}} = -1$ on the $k_i = \pi$ plane, which leads to Kramer-like degeneracy and the formation of the nodal surface. We noticed that the nodal surface phonons in the BaSi$_2$ material with space group $P4{_3}32$ have also been mentioned in reference~\cite{xie2021three}.

\section{Unique Surface Fermi Arc States}
\begin{figure}[htb]
	\includegraphics[width=8.5cm]{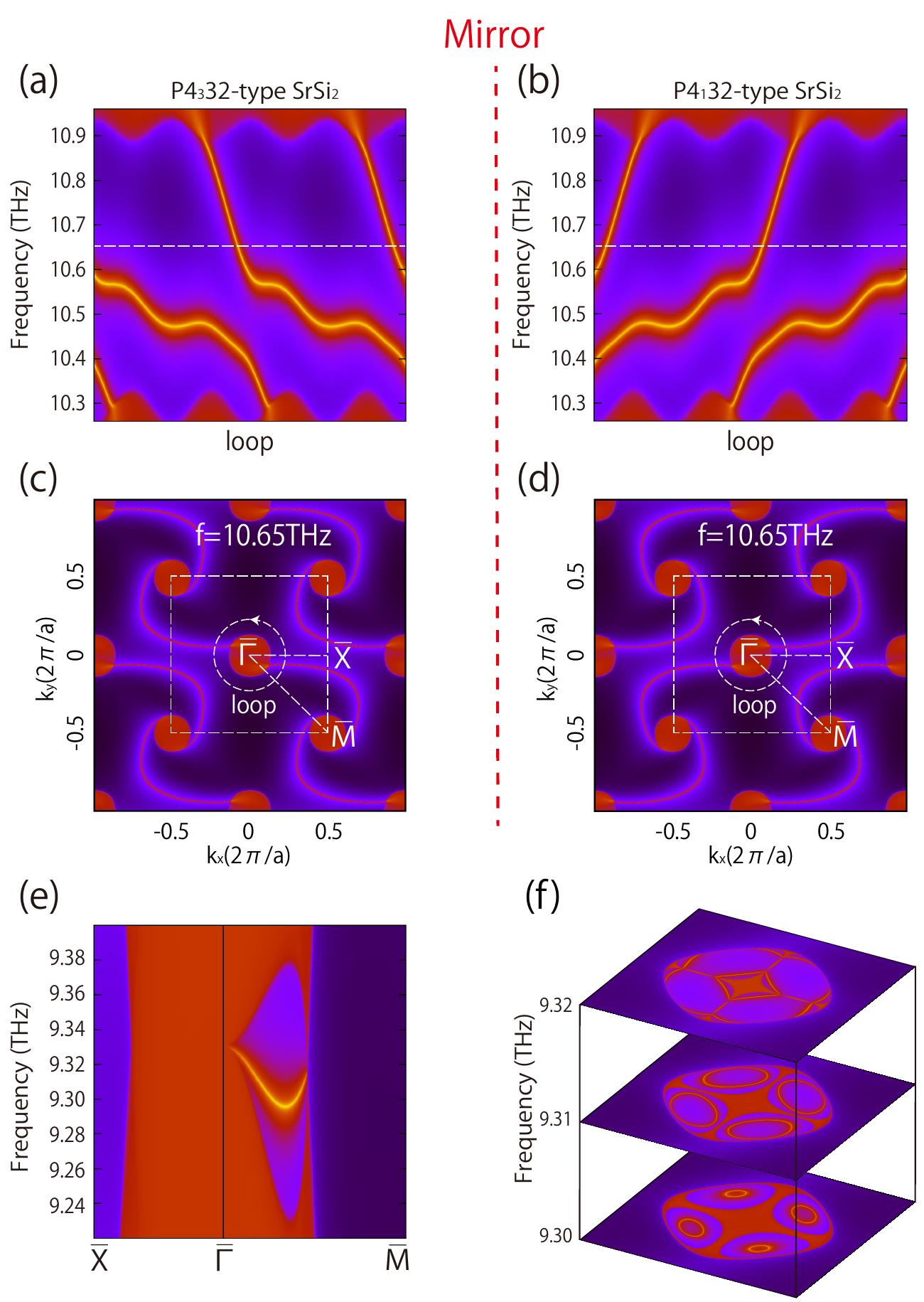}
	\caption{Surface states and Fermi arcs of the (001) surface for the C-2 TP and C-2 DP: (a) and (c) are for the SrSi$_2$ with SG No. 212, and (b) and (d) are for the SrSi$_2$ with SG No. 213. (e) and (f) are surface states and Fermi arcs of the (001) surface for the C-2 WP and C-1 WP for the SrSi$_2$ with SG No. 212. 
		\label{fig4}}
\end{figure}

Topological quasiparticles typically exhibit surface states. The spectra of the (001) surface for the SrSi$_2$ material are shown in Fig.~\ref{fig4}. Figures.~\ref{fig4} (a) and (c) depict the surface states and Fermi arcs at 10.65 THz for the SrSi$_2$ material with SG No. 212, while Figs.~\ref{fig4} (b) and (d) correspond to the SrSi$_2$ material with SG No. 213. These surface states arise from the C-2 TP and C-2 DP. Fang et al. demonstrated that the surface states of Weyl points are equivalent to a helicoid~\cite{fang2016topological}. Here, since the surface states arise from the C-2 TP and C-2 DP, two surface sheets encircle these points, resulting in double-helicoid surface states, as illustrated in Figs.~\ref{fig4}(a) and (b). Furthermore, two large topological surface states emerge from $\bar{\Gamma}$ and $\bar{M}$. Here, the $\bar{\Gamma}$ point represents the projection of the C-2 TP at $\Gamma$, and $\bar{M}$ represents the projection of the C-2 DP at $R$. Additionally, large Fermi arcs connect the points $\bar{\Gamma}$ and $\bar{M}$, spanning the entire surface BZ, as shown in Figs.~\ref{fig4}(c) and (d). Notably, it is clearly observed that the surface states of the two enantiomers exhibit opposite chirality. Therefore, the crystal structural chirality also induces the chirality of topological surface states. 

In Figs.~\ref{fig4} (e) and (f), we also present the surface state and Fermi arc for the C-2 WP and C-1 WP of SrSi$_2$ material with SG No. 212 (surface state and Fermi arc for the SrSi$_2$ with SG No. 213 are similar). Due to the presence of six C-2 WPs with charge +2 and twelve C-1 WPs with charge -1 in the BZ, the no-go theorem is satisfied~\cite{nielsen1981absence}. Therefore, the surface states should originate from the C-2 WPs and connect to the C-1 WPs. The surface states of the (001) surface of SrSi$_2$ with SG No. 212 are shown in the Figs.~\ref{fig4} (e) and (f). One can identify the topologically protected surface arcs terminating at the projections of the C-2 WPs and C-1 WPs. It is worth noting that the surface arcs shown here are distinctly different from the surface states of conventional Weyl points, as these arcs connect Weyl points with unequal charge, resulting in a three-terminal Weyl complex with two surface arcs. Additionally, due to the presence of $C_4$ symmetry, the arc states on the (001) plane exhibit distinct tetragonal symmetry.

\section{Discussion and Conclusion}

In this study, we explore the relationship between the chirality of crystal structures and the chirality of topological quasiparticles, demonstrating that enantiomers with opposite chirality produce opposite chiral topological charges. Utilizing first-principles calculations and theoretical analysis, we show that various topological quasiparticles—including charge-2 triple points, charge-2 Dirac points, charge-2 Weyl points, and charge-1 Weyl points—can be realized in the phonon spectra of chiral crystal materials like SrSi$_2$ and BaSi$_2$. Our findings indicate that these topological quasiparticles in the enantiomers exhibit distinct chiral topological charges and unique topological surface states. Moreover, we identify similar topological quasiparticles in the phonon spectra of other chiral crystal materials belonging to space group No. 213, such as Nb$_3$Al$_2$N, Nb$_3$Al$_2$C, Ta$_3$Al$_2$C, and Cs$_3$P$_6$N$_{11}$ (see the SM~\cite{SM}).

It should be noted that the conclusions of our paper apply not only to enantiomeric space groups 212 and 213, but also to other enantiomeric space groups. For example, Te materials with chiral enantiomeric space groups 152 and 154 also exhibit Weyl points with opposite chirality charges due to the chirality of their structures~\cite{zhang2023weyl, pan2024intrinsic}. Since 22 chiral space groups form 11 pairs of enantiomers in three-dimensional (3D) systems~\cite{fecher2022chirality}, it is also worth further studying the chiral topological quasiparticles (including topological electrons and topological phonons) in other enantiomeric structures. Experimentally, the chirality of crystal structures can be measured using various methods, such as the continuous chirality measure~\cite{zabrodsky1992continuous,zabrodsky1995continuous}, the Hausdorff distance~\cite{fecher2022chirality,buda1992hausdorff},
the pseudo-angular momentum~\cite{streib2021difference}, and the helicity~\cite{gomez2024pros,bousquet2024structural}. Meanwhile, the chirality of topological phononic quasiparticles in
chiral crystals can be probed by Raman scattering~\cite{zhang2023weyl}. Our results offer new perspectives and a promising platform for exploring the intriguing physics associated with chiral crystals, including phenomena such as the chirality-dependent nonlinear Hall effect and controllable orbital angular momentum monopoles~\cite{pan2024intrinsic,yen2024controllable}.

\bigskip
\begin{acknowledgements}
	This work is supported by the National Natural Science Foundation of China (Grant No. 12204378). H. H. acknowledges the support of the National Key R\&D Program of China (Grant No. 2021YFA1401600) and the National Natural Science Foundation of China (Grant No. 12074006).
\end{acknowledgements}

%

\end{document}